# Strong interrelationship between anomalous electric-field induced lattice strain along non-polar direction and domain reorientation in pseudorhombohedral piezoelectric ceramic BiScO$_3$-PbTiO$_3$

Lalitha KV[1], Chris M. Fancher[2], Jacob L. Jones[2] and Rajeev Ranjan[1*]

[1]Department of Materials Engineering, Indian Institute of Science, Bangalore 560012, India
[2]Department of Materials Science and Engineering, North Carolina State University, Raleigh, NC 27695, USA

**Abstract**

The lattice strain and domain switching behavior of a pseudorhombohedral composition of the piezoelectric ceramic 0.40BiScO$_3$ – 0.60PbTiO$_3$ was investigated as a function of cyclic field and grain orientation by *in situ* X-ray diffraction during application of electric fields. The electric field induced 200 lattice strain was measured to be five times larger than the 111 lattice strain. In difference with some of the interpretations in the past that considered this phenomenon of anomalous lattice strain along non-polar direction to be a manifestation of polarization rotation within the unit cell, the anomalous 200 lattice strain is shown to be intimately associated with the reorientation of the 111 domains in the dense polycrystalline ceramic.

* rajeev@materials.iisc.ernt.in

Ferroelectric alloy systems exhibiting morphotropic phase boundary (MPB), such as lead-zirconate-titanate (PZT), BiScO$_3$-PbTiO$_3$ (BSPT) are of great scientific and technological importance because of their use as pressor-sensors and actuators[1-5]. Though such piezoelectric materials are being extensively used, the mechanisms leading to the anomalous piezoelectric response is still a subject matter of considerable debate. In general, there are three important contributing mechanisms with regard to the overall piezoelectric response in MPB ferroelectrics: (i) lattice strain, (ii) domain wall displacement, and (iii) interferroelectric phase transformation. While one viewpoint attributes the anomalous piezoelectric response of the MPB compositions to polarization rotation along low energy pathways in the unit cell[6], another view point based on martensitic theory considers enhanced mobility of nano domains as the origin of anomalous piezoresponse[7,8]. Since the direction of spontaneous polarization differs in different crystallographic phases of a ferroelectric material, field induced interferroelectric phase transformation can be argued as evidence of polarization rotation/switching[9]. This aspect was highlighted by Lalitha *et al.* who reported a decrease in volume-fraction of the coexisting monoclinic phase as compared to the tetragonal phase after poling of the MPB composition of BSPT[10]. Jones *et al.*, on the other hand, have reported dominant contribution of domain wall motion in the coexisting monoclinic phase of this system[11]. Earlier, the polarization rotation model was supported by Guo et al[12] in PZT ceramic based on the observation of anomalous field induced 200 lattice strain in pseudorhombohedral/monoclinic phase. It may be noted that the polar direction in rhombohedral ferroelectric perovskite is [111], and the large lattice strain along the non-polar 200 direction was interpreted as a manifestation of the enhanced tendency of the polarization vector to rotate away from the polar direction which was in accordance with the prediction of Du et al[13] for single crystals. Crystallographic studies have therefore reported the polarization rotation mechanism to manifest in two different ways: (i) field induced inter-ferroelectric transformation and (ii) anomalous lattice strain along non polar directions. In the present work, we have carried out a combined analysis of lattice strain and domain reorientation in a pseudorhombohedral/monoclinic composition of the BSPT ($x = 0.40$) in close proximity to, but not directly on, the MPB. The single phase nature of this composition enabled simultaneous quantification of the field induce domain reorientation and lattice strains, reliably. Our results show that the anomalous lattice strain along non-polar direction is not an independent intrinsic mechanism, but is intimately related to, and governed by elastic strain induced by reorientation of the non-180° domains.

Samples of $x$BiScO$_3$-(1-$x$)PbTiO$_3$ $x = 0.40$ (BS40) were prepared by the conventional solid, the details of which are given in [10, 14]. Sintered pellets were cut into bars (10 x 1 x 1 mm) and electric fields were applied across parallel 10 mm x 1 mm faces. *In situ* x-ray diffraction was carried out at the Advanced Photon Source at Argonne National laboratory in transmission geometry that ensures that the measured diffraction data probes the bulk response of the specimen. A monochromatic X-ray beam of wavelength 0.11165Å and size 500 x 500 μm was used for the diffraction experiments. Diffraction data was measured during application of bipolar triangular electric fields (300 s period) of maximum amplitude between 2.5 and 4.5 kV/mm, in increments of 0.5 kV/mm. Measured diffraction data was recorded every 5 s using a 2D detector wherein the circular Debye rings correspond to different *hkl* reflections. The diffraction images were divided into 24 azimuthal sectors (ψ) of 15° widths, with the azimuthal sector most closely oriented to the direction of applied electric field defined as ψ = 0°, (Fig. 1). The diffracted intensities as a function of 2θ were obtained by reducing each azimuthal sector using the software FIT2D[15].

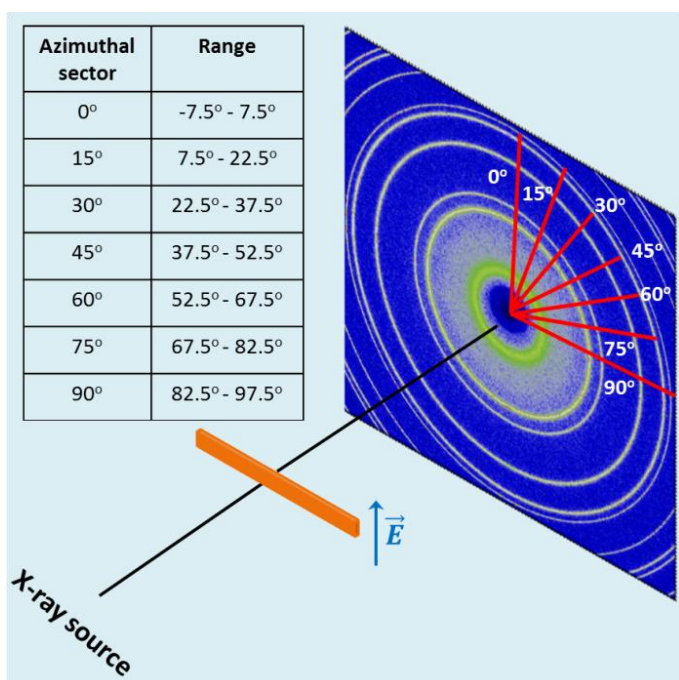

**Fig. 1** Experimental geometry for *in situ* x-ray diffraction in transmission mode. The 2D image of the Debye rings corresponding to different (*hkl*) diffracted beams is divided into 24 azimuthal sectors of 15° widths. The azimuthal sectors are marked for the first quadrant on the image and the corresponding ranges are shown in the table on the left.

Prior work used high-resolution XRD to demonstrate that BS40 exhibits a monoclinic (*Cm*) symmetry[5,10]. Figure 1 shows that the 111 and 200 reflections are consistent with a rhombohedral symmetry within the resolution limit of the present instrumental setup. In the present, work, the structure is considered as pseudorhombohedral (pr), and the *hkl* reflections are indexed using the pseudorhombohedral cell. The evolution of the 111 and 200 reflections as a function of electric field for ψ = 0° and also as function of azimuth angle for the maximum field (E=2.5 kV/mm) are compared in Figure 2. Application of an electric field of magnitude > 1.5 kV/mm induces ferroelectric/ferroelastic reorientation as evidenced by an intensity interchange between the 111 and $11\bar{1}$ reflections, and a shift in the position of the 200 reflection. In contrast to the behavior of the 200 reflection, the peak position of the 111 reflections are independent of applied field strength.

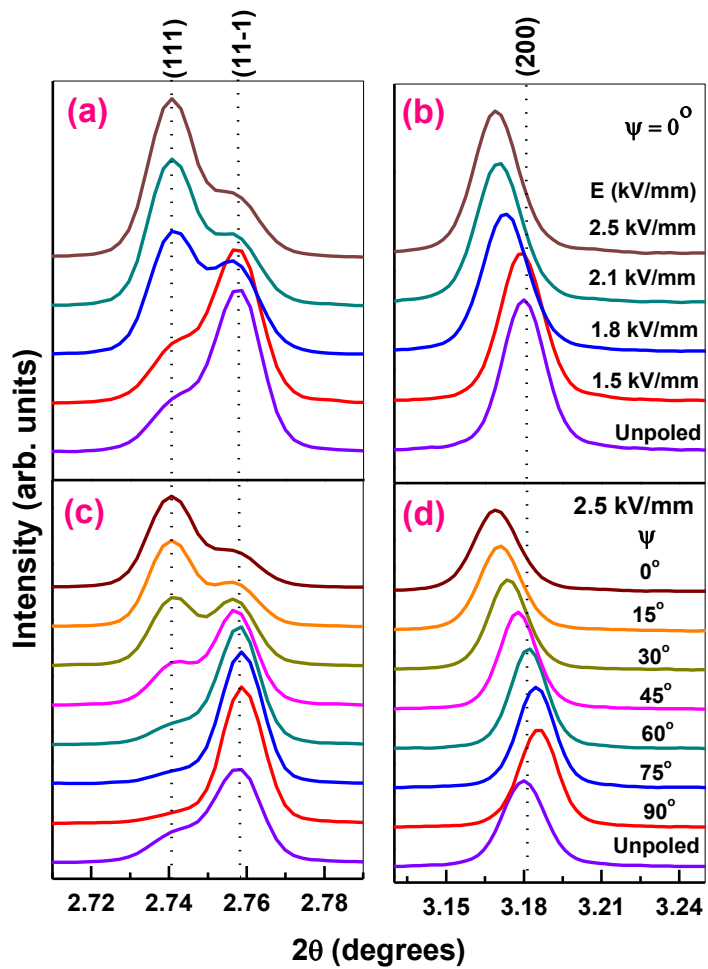

**Fig 2**. X-ray diffraction spectra of (111)/(11-1) and (200) pseudorhombohedral Bragg profiles of BS40 showing shift in peak positions as a function of electric field for $\psi = 0°$ (a-b) and orientation with respect to the direction of electric field (c-d) for an applied electric field amplitude of 2.5kV/mm.

Measured diffraction spectra were analyzed using single peak fitting to elucidate additional information about the field-induced ferroelectric/ferroelastic reorientation and lattice strain.[16] By assuming a pseudo-rhombohedral symmetry, the split 111 Bragg peaks can be fit with two profiles, and the 200 can be fit with a single profile as shown in Fig. 3.

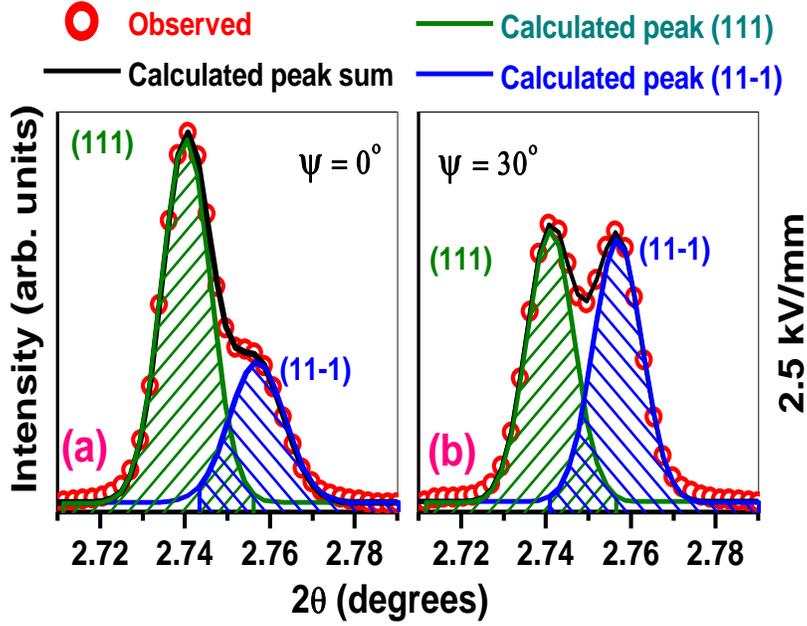

**Fig. 3** Observed (open circles) and fitted diffraction peaks of the $\{111\}_{pc}$ reflection of BS40 for electric field amplitude of 2.5 kV/mm for orientation $\psi$ (a) 0° (parallel) and (b) 30° to the direction of electric field. The observed data is fit using two Gaussian profile functions and the integration of the individual peaks is terminated beyond the peak position of the adjacent peak as shown by the shaded regions.

The extracted intensities of the 111 and $11\bar{1}$ reflections can be used to determine the volume fraction of domains that have been reoriented: [17, 18]

$$\eta_{111} = \frac{\frac{I_{111}}{I'_{111}}}{\frac{I_{111}}{I'_{111}} + 3\frac{I_{11\bar{1}}}{I'_{11\bar{1}}}} - \frac{1}{4}$$

where, $I_{111}$ and $I_{11\bar{1}}$ are the integrated intensities of the 111 and $11\bar{1}$ reflections for a poled state, respectively. $I'_{111}$ and $I'_{11\bar{1}}$ are the integrated intensities of the 111 and $11\bar{1}$ reflections in the unpoled state, respectively. A value of $\eta_{111} = 0.75$ corresponds to a case of complete domain reorientation. Figure 4a shows the evolution of $\eta_{111}$ for the initial application of electric field (up to 2.5 kV/mm) to an unpoled sample for $\psi = 0^0$. The initial field application induces a sharp increase in $\eta_{111}$ beyond a threshold field of 1.5 kV/mm.

A maximum in $\eta_{111}$ of 0.28 is induced for an applied field of 2.5 kV/mm. As the applied electric field is decreased from 2.5 kV/mm to 0 kV/mm, $\eta_{111}$ decreases by 11% to 0.25 and further decreases until the negative coercive field (-1.5 kV/mm), then sharply increases again.

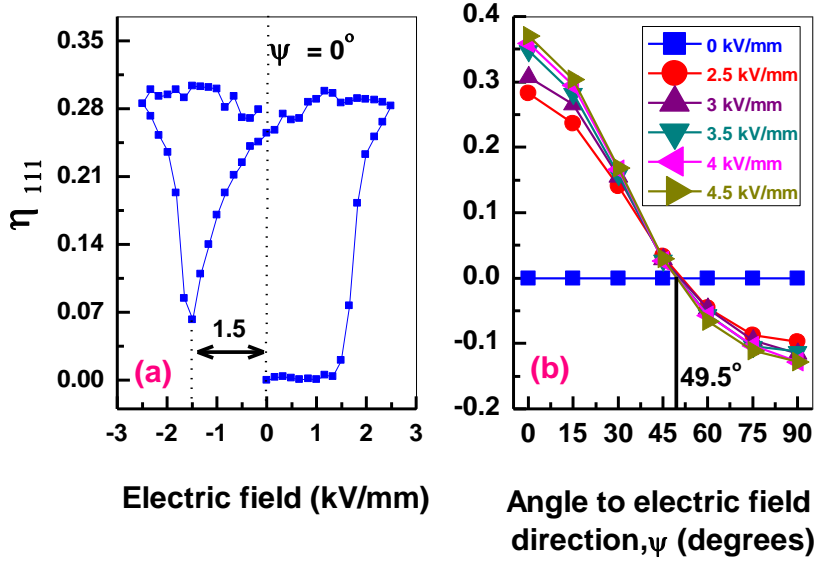

Fig 4. Variation of (a) $\eta_{111}$ ($\psi = 0°$) during application of bipolar electric fields with a 2.5 kV/mm amplitude. (b) Angular dependence of $\eta_{111}$ for various applied field amplitudes shows an increase in $\eta_{111}$ for angles that are closely aligned to the direction of the applied fields.

The slight reduction in a minimum in $\eta_{111}$ (0.06) is observed at 1.5 kV/mm, suggesting that the coercive field is ~1.5 kV/mm. The obtained coercive field differs slightly with the coercive field obtained from macroscopic measurement (1.8 kV/mm). The difference in the coercive field measured from this work is likely a result of the longer waveform period. This result is consistent with the known time and frequency dependence of domain switching in ferroelectric systems[19-21]. Figure 4b shows the orientation dependence of $\eta_{111}$. This quantity is maximum for $\psi = 0°$ and decreases with increasing $\psi$. For $\psi > 45°$ $\eta_{111}$ is negative. Except for the difference in the absolute values, this trend remains the same for different magnitude of the electric field amplitudes. From the trend, it can be anticipated that $\eta_{111} = 0$ at $\psi \sim 49°$. This angle is close to 45° predicted for a rhombohedral symmetry[17]. $\eta_{111}=0$ near $\psi \sim 49°$ implies that domain switching is energetically unfavorable around this angle.

The peak positions extracted from single peak fitting were used to quantify the field induced lattice strains $\varepsilon_{111}$ and $\varepsilon_{200}$. Figure 5a shows the lattice strain $\varepsilon_{200}$ parallel to the field during the application of bipolar triangular electric fields with amplitude of 2.5 KV/mm. A maximum $\varepsilon_{200}$ of 0.33% was induced with a remanent strain of 0.22%.

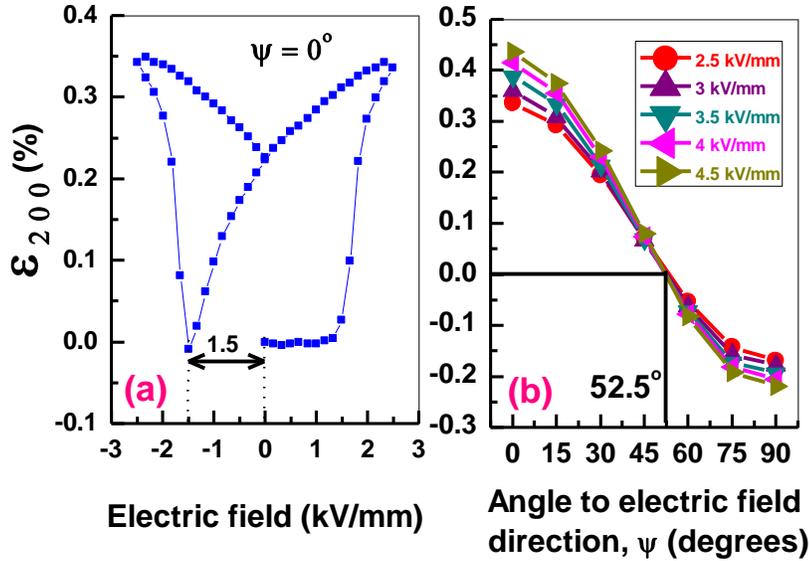

Fig 5. (a) Induced 200 lattice strain during application of bipolar electric fields with a 2.5 kV/mm amplitude. (b) Angular dependence of the induced 200 lattice strain as a function of electric field amplitude and orientation with respect to the electric field direction ($\psi$).

The field dependence of $\varepsilon_{200}$ is similar to a butterfly loop observed in macroscopic strain-field measurements. The field induced lattice strain obtained from the present study is 0.34% at 2.5 kV/mm, which is roughly twice that obtained from *in situ* field dependent diffraction studies on coarse grained La(2 mol%)-modified rhombohedral PZT[22]. Additionally, the field-dependent change in $\varepsilon_{200}$ (Fig. 5a) mimics the hysteretic behavior observed in $\eta_{111}$ (Figure 4a) and the orientation dependence of $\varepsilon_{200}$ (Fig. 5b) resembles the angular dependence in $\eta_{111}$ (Figure 4b). The field at which the strain is zero (-1.5 kV/mm) coincides with the field where a minimum in $\eta_{111}$ was observed. The one-to-one correspondence between $\varepsilon_{200}$ and $\eta_{111}$ suggest that the induced lattice strain and non-180° domain reorientation are strongly coupled. Hall *et al.*[23] suggested that the $\varepsilon_{200}$ lattice strain is characteristic of the intergranular stresses resulting from grain-to-grain interactions and predicted a linear dependence of the strain with $\sin^2\psi$.[24,25] We found this relationship holds true between $\varepsilon_{200}$ and $\sin^2\psi$ (Figure 6) for our specimen.

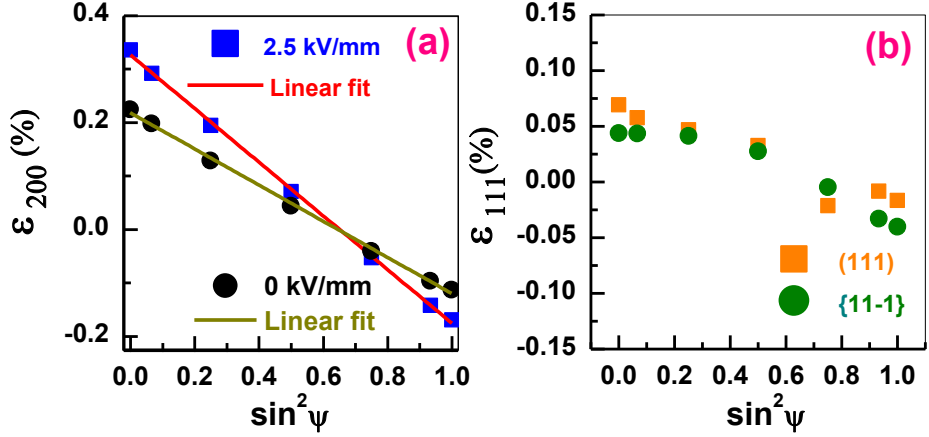

Fig. 6 Variation of (a) $\varepsilon_{200}$ at E = 2.5 kV/mm (square) and after removal of the field (filled circles) and (b) $\varepsilon_{111}$ with $\sin^2\psi$ at 2.5 kV/mm.

The linear behavior of $\varepsilon_{200}$ can be used to evaluate the transverse and parallel strains[27] using

$$(\varepsilon'_{33})_{\varphi\psi} = \frac{d_{\varphi\psi} - d_0}{d_0}$$
$$= \{\varepsilon_{11} \cos^2\varphi \sin^2\psi + \varepsilon_{12} \sin2\varphi \sin^2\psi + \varepsilon_{22} \sin^2\varphi \sin^2\psi + \varepsilon_{33} \cos^2\psi$$
$$+ \varepsilon_{13} \cos\varphi \sin2\psi + \varepsilon_{23} \sin\varphi \sin2\psi\} \quad (1)$$

wherein, $(\varepsilon'_{33})_{\varphi\psi}$ is the projected normal strain. $\varphi$ is the angle between a fixed direction in the plane of the sample and the projection of the normal of the diffracting plane in the plane of the sample. $\psi$ is the angle between the normal of the sample, N and the normal of the diffracting plane, $N_d$ (bisecting the incident and diffracted beams) as shown in Fig. 7. In the present case, $\psi$ is taken as the angle between the scattering vector and electric field direction. The electric field is applied normal to the sample. $d_{\varphi\psi}$ is the inter-planar spacing of planes measured at angles defined by $\varphi$ and $\psi$. $\varepsilon_{ij}$, i = j represents the principle strains and i ≠ j represents the shear strains. Since $\varepsilon_{200}$ is a linear function of $\sin^2\psi$, the shear strains must be negligible, or $\varepsilon_{13} = \varepsilon_{23} = 0$. Assuming the sample exhibits a fiber symmetry,[26] $\varepsilon_{12} = 0$ and $\varepsilon_{11} = \varepsilon_{22}$ and eq. (1) simplifies to

$$(\varepsilon'_{33})_\Psi = \{2\varepsilon_{22} - \varepsilon_{33}\}sin^2\Psi + \varepsilon_{33} \qquad (2)$$

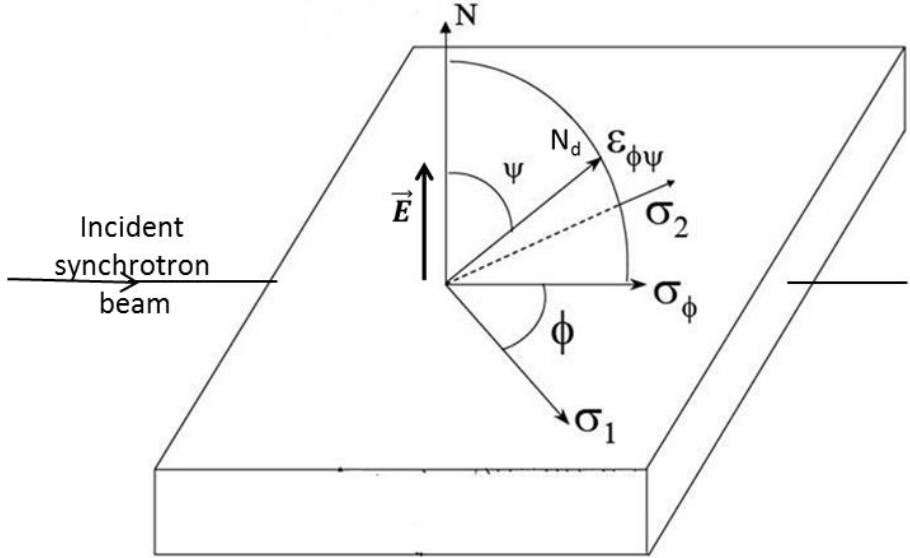

Fig. 7 Schematic illustrating diffraction planes parallel to the surface and at an angle φ and Ψ.

A linear fit of the experimental data yields the percentage transverse strains $\varepsilon_{11} = \varepsilon_{22} = -0.175 \pm 0.003$ and percentage parallel strain $\varepsilon_{33} = 0.327 \pm 0.004$. Upon application of electric field, 200 lattice spacings elongate parallel to the electric field and contract in directions perpendicular to the direction of the electric field. It is interesting to note that $\varepsilon_{200}$ remains substantially large ($\varepsilon_{33} = 0.22$ along the field direction) and follows a linear dependence with $sin^2 \Psi$ even after the removal of the electric field (Figure 5a). In contrast to the MPB compositions for which field induced pseudo-rhombohedral to tetragonal transformation has been reported in the past[10,] the diffraction pattern of the present composition did not show any signature of additional peak to suggest a field induced phase transformation. For example, if the tetragonal phase were to appear, two peaks corresponding to 002 and 200 of the tetragonal phase are expected to appear to 200 rhombohedral peak. However, no such symptom could be detected in Fig. 1b and

1d even at the highest field. In fact, there is not even a noticeable broadening of the existing 200 rhombohedral peak to indicate a possible phase transformation. Absence of phase transformation in this composition was also reported earlier in a diffraction study of powder obtained from a poled specimen.[10] In the absence of field induced phase transformation, polarization rotation as a dominant mechanism is less likely for this composition[9,27]. The large residual $\varepsilon_{200}$ lattice strain is primarily due to stress generated by domains that have been reoriented after application of the field. The same scenario persists when the field in on. Hence, although the anomalous 200 lattice strain observed in the present study is similar to what was reported earlier for rhombohedral/monoclinic PZT[12], the interpretations are distinctly different. Our results suggest that the large 200 lattice strain cannot be exclusively attributed to the polarization rotation, but has its origin primarily in the elastic strain resulting from lattice mismatch due to reorientation of domains in differently oriented grains in a polycrystalline specimen.

To summarize, we carried out *in situ* electric field X-ray diffraction studies on a psedorhombohedral composition (x=0.40) close to the MPB of the ferroelectric alloy system $xBiScO_3$-$(1-x)PbTiO_3$. Similar to what has been reported for pseudo-rhombohedral PZT, this system was also found to exhibit anomalous 200 lattice strain. Analysis of the grain orientation dependence of this anomalous strain during application of electric field, and also after its removal, revealed that the anomalous 200 lattice strain is elastic in nature, and arise due to stress field resulting from field induced domain reorientation. Our results therefore does not seem to favor a direct correspondence between polarization rotation and anomalous lattice strain in non-polar direction, as has been considered in the past, and offers a suitable background for better understanding the dominant contributing mechanisms associated with the anomalous piezoelectric response in structurally and microstructurally more complex MPB compositions.

RR thanks the Science and Engineering Board of the Department of Science and Technology for financial assistance (Grant number: SERB/F/5046/2013-14).

[1] R. Eitel, C. Randall, T. Shrout, P. Rehrig, W. Hackenberger, and S.-E. Park, Jpn. J. Appl. Phys. **40**, 5999 (2001).

[2] R.E. Eitel, C.A. Randall, T.R. Shrout, and S.-E. Park, Jpn. J. Appl. Phys. **41**, 2099 (2002).

[3] J. Chaigneau, J.M. Kiat, C. Malibert, and C. Bogicevic, Phys. Rev. B **76**, 094111 (2007).

[4] B. Kim, P. Tong, D. Kwon, J.M.S. Park, and B.G. Kim, J. Appl. Phys. **105**, 114101 (2009).

[5] K. Datta, D. Walker, and P.A. Thomas, Phys. Rev. B **82**, 144108 (2010).

[6] H. Fu and R.E. Cohen, Nature **403**, 281 (2000).

[7] Y.M. Jin, Y.U. Wang, A.G. Khachaturyan, J.F. Li, and D. Viehland, Phys. Rev. Lett. **91**, 197601 (2003).

[8] Y.U. Wang, Phys. Rev. B **76**, 024108 (2007).

[9] M. Davis, D. Damjanovic, and N. Setter, Phys. Rev. B **73**, 014115 (2006).

[10] Lalitha K. V., A.N. Fitch, and R. Ranjan, Phys. Rev. B **87**, 064106 (2013).

[11] J.L. Jones, E. Aksel, G. Tutuncu, T.-M. Usher, J. Chen, X. Xing, and A.J. Studer, Phys. Rev. B **86**, 024104 (2012).

[12] R. Guo, L. E.Cross, S.-E. Park, B. Noheda, D. E. Cox, and G. Shirane Phys Rev Lett **84** 5423 (2000)

[13] X. Du, U. Belegundu, and K. Uchino, Jpn. J. Appl. Phys. **36**, 5580 (1997)

[14] Lalitha.K.V., A.K. Kalyani, and R. Ranjan, Phys. Rev. B **90**, 224107 (2014).

[15] A.P. Hammersley, S.O. Svensson, M. Hanfland, A.N. Fitch, and D. Hausermann, High Press. Res. **14**, 235 (1996).

[16] G. Esteves, C.M. Fancher and J.L. Jones, J. Mater. Res., 30, 340-356. (2015)

[17] J.L. Jones, M. Hoffman, and K.J. Bowman, J. Appl. Phys. **98**, 024115 (2005).

[18] A. Pramanick, D. Damjanovic, J.E. Daniels, J.C. Nino, and J.L. Jones, J. Am. Ceram. Soc. **94**, 293 (2011

[19] C.F. Pulvari and W. Kuebler, J. Appl. Phys. **29**, 1315 (1958).

[20] M.H. Lente, A. Picinin, J.P. Rino, and J.A. Eiras, J. Appl. Phys. **95**, 2646 (2004).

[21] W. Li, Z. Chen, and O. Auciello, J. Phys. Appl. Phys. **44**, 105404 (2011).

[22] M.J. Hoffmann, M. Hammer, A. Endriss, and D.C. Lupascu, Acta Mater. **49**, 1301 (2001)

[23] D.A. Hall, A. Steuwer, B. Cherdhirunkorn, T. Mori, and P.J. Withers, J. Appl. Phys. **96**, 4245 (2004).

[24] D.A. Hall, A. Steuwer, B. Cherdhirunkorn, P.J. Withers, and T. Mori, J. Mech. Phys. Solids **53**, 249 (2005).

[25] D.A. Hall, A. Steuwer, B. Cherdhirunkorn, P.J. Withers, and T. Mori, Ceram. Int. **34**, 679 (2008).

[26] D.I.C. Noyan and P.J.B. Cohen, in *Residual Stress* (Springer New York, 1987), pp. 117–163.

[27] M. Davis, M. Budimir, D. Damjanovic, and N. Setter, J. Appl. Phys. **101**, 054112 (2007).